# Charge transport in ion-gated mono-, bi-, and trilayer MoS$_2$ field effect transistors

*Leiqiang Chu\*[1,2], Hennrik Schmidt\*[1,2], Jiang Pu[4], Shunfeng Wang[1,2], Barbaros Özyilmaz[1,2], Taishi Takenobu[4,5,6], Goki Eda[1,2,3,#]*

[1]Graphene Research Centre, National University of Singapore, 6 Science Drive 2, Singapore 117546

[2]Department of Physics, National University of Singapore, 2 Science Drive 3, Singapore 117542

[3]Department of Chemistry, National University of Singapore, 3 Science Drive 3, Singapore 117543

[4]Department of Advanced Science and Engineering, Waseda University, Tokyo 169-8555, Japan

[5]Department of Applied Physics, Waseda University, Tokyo 169-8555, Japan

[6]Kagami Memorial Laboratory for Material Science and Technology, Waseda University, Tokyo 169-0051, Japan

\* These authors contributed equally to this work.

# Abstract

Charge transport in $MoS_2$ in the low carrier density regime is dominated by trap states and band edge disorder. The intrinsic transport properties of $MoS_2$ emerge in the high density regime where conduction occurs via extended states. Here, we investigate the transport properties of mechanically exfoliated mono-, bi-, and trilayer $MoS_2$ sheets over a wide range of carrier densities realized by a combination of ion gel top gate and $SiO_2$ back gate which allows us to achieve high charge carrier ($>10^{13}$) density. We discuss the gating properties of the devices as a function of layer thickness and demonstrate resistivities of as low as 1 kΩ for monolayer and 420 Ω for bilayer devices at 10 K. We show that from the capacitive coupling of the two gates, quantum capacitance can be roughly estimated to be on the order of 1 μF/cm$^2$ for all devices studied. Temperature dependence of the carrier mobility in the high density regime indicates that short-range scatterers limit charge transport at low temperatures.

Two-dimensional (2D) crystals of layered transition metal dichacogenides (TMD) have gained significant interest due to their unique physical properties.[1, 2] Molybdenum disulfide ($MoS_2$), a semiconducting compound traditionally known for its lubricating properties, has been intensely investigated in its 2D crystalline form due to its stability and accessibility. Field effect transistors of monolayer $MoS_2$ have shown remarkable characteristics with low off-current due to excellent gate electrostatics[3]. Recent studies on the implementation of $MoS_2$ and other semiconducting TMDs into integrated circuits[4-7] and optoelectronic devices[8-10] highlight its potential in future applications.

Electrostatically controlled switching and charge transport in mono- and few-layer $MoS_2$ devices are strongly influenced by various effects such as metal contacts[11, 12], interface traps[13, 14], charge impurities[3, 13, 15], dielectric environment[3, 4, 6, 16, 17], and structural defects[18-22] in the material. The contribution of these effects varies with the gate bias in a complex manner. At low gate biases near the sub-threshold region, the resistance is strongly dominated by Schottky barriers of the contacts[11] and hopping-transport.[13, 20, 22-24] Band edge disorder and mid-gap states originating from defects, charge traps at $MoS_2/SiO_2$ interface, and surface adsorbates[13, 14] play a persistent role in this low charge density regime. At higher charge densities, the transition to metallic like conduction indicative for band transport has been observed at a resistance of $h/e^2$.[17, 20, 25, 26] In this regime, charged impurities, defects, and surface-optical phonons[27] limit the charge carrier mobility. All recent studies show that low temperature mobility of mono- and bilayer $MoS_2$ falls substantially below theoretically predicted values[28, 29], suggesting that there is significant room for improvement in the device performance.

The use of high-κ dielectrics, such as $HfO_2$[3, 17, 30] and $Al_2O_3$[16, 31], as an encapsulating layer and gate barrier has been reported to be effective in enhancing carrier mobility, potentially due to dielectric screening of charged impurities and suppression of homopolar optical phonon modes.[17, 32] The high-κ dielectric in combination with a topgate electrode also allows access to the high density regime where transport is band-like and the intrinsic transport properties of the material can be investigated. In order to achieve even higher charge densities, ionic gating has been used in different 2D materials[33-38]. Previous studies using polymer electrolyte[39], ionic liquid[26, 37, 40, 41], and ion gels[42] have demonstrated effective switching and high doping levels of up to ~$10^{14}$ cm$^{-2}$ in few-layer $MoS_2$ devices[41, 42] to realize flexible transistors[42], stable p-n junctions[34, 40, 43], light emitting devices[31, 34, 43] and superconductivity.[26]

In this article, we report on the low temperature transport characteristics of mono- , bi- and trilayer $MoS_2$ devices in the high doping regime using ion gel gating. We show that large capacitive coupling of the ion gel in conjunction with additional electrostatic control by the back gate allows systematic investigation of charge transport over a wide range of charge carrier densities. Resistivities as low as 1 kΩ and 420 Ω are realized in highly doped mono- and bilayer $MoS_2$ at low temperature, respectively. From the capacitive coupling of the two gates, we estimate the quantum capacitance, which is a measure of the density of states in these materials. Further analysis of the temperature- and density-dependent field effect mobility reveals that short-range scatterers severely limit carrier mobility at low temperatures in all devices.

Atomically thin flakes of $MoS_2$ sheets were mechanically exfoliated from bulk crystals (SPI supplies) and subsequently deposited onto silicon substrate with 300 nm thermal $SiO_2$. The flake thickness is estimated by optical contrast and then further confirmed by the peak separation between the $A_{1g}$ and $E_{2g}^1$ peaks in the Raman spectrum.[44] In agreement with previous studies [13], the peak separation was found to be 18.3 cm$^{-1}$, 21.2 cm$^{-1}$, and 23.1 cm$^{-1}$ for monolayer (1L), bilayer (2L), and trilayer (3L) samples, respectively. Pure gold contacts (50 nm) for source, drain, and side gate electrodes were fabricated using standard electron beam lithography and thermal evaporation. After lift-off, all devices were annealed in nitrogen atmosphere at 200 °C for 2 hours. The ion gel solution was prepared by mixing the polymer PS-PMMA-PS and the ionic liquid EMIM-TFSI (Figure 1c) into an ethyl propionate solvent (weight ratio of polymer: ionic liquid: solvent=0.7: 9.3:90).[42] A film of this solution was spin-coated onto the devices and dehydrated in nitrogen gas for an hour to remove molecular moisture. Figure 1a shows optical images of three devices of different thickness covered with a thin ion gel film.

Transport measurements were performed in vacuum inside a Helium 4 cryostat with variable temperature insert. Figure 1b shows the transfer curves of mono-, bi-, and trilayer samples while varying the voltage applied to the ionic gate at room temperature, demonstrating the wide tunability of the Fermi level achieved by ion gel. For bilayer and trilayer devices, both electron and hole branches are observed at positive and negative gate voltages, respectively. The monolayer device only shows the electron branch in the top gate bias window studied due to its large band gap. Assuming that the variation of the electrostatic potential is negligible, the bandgap of bilayer and trilayer samples can be roughly estimated to be 1.63 eV and 1.32 eV, which is in good accordance with the optical gap.[45] Figure 1d displays the temperature dependent transconductance of a bilayer device, obtained by sweeping the top gate voltage (4.5 mV/s) at a fixed temperature. Below T~210 K, the top gate modulation becomes negligible, suggesting the freezing of the ions in the gel matrix. This freeze out allows us to achieve stable high doping in the channel and fine control of the carrier density by the back gate below this temperature.

The dual gating behavior of the devices was studied by cooling them below the critical freezing point at a fixed top-gate voltage and measuring the transfer characteristics by back gate sweeps at various temperatures. To study a device over a wide range of charge densities, the sample was brought to room temperature and the topgate voltage was set to a desired value and held until the ion motion was negligible. The sample was then cooled below the freezing point for measurements. The above procedure was repeated with a different topgate biases. Figures 2a-c show the backgate transfer characteristic of the devices at different top gate biases. At negative $V_{tg}$, a transition from the insulating to the conducting state can be observed with on-off ratios of > $10^5$ for all devices. The off-state conductance was limited by the leakage current while the on-state conductance was in the same order of magnitude for the three cases. It is worth noting that all samples exhibit low sheet resistivity (< 1 kΩ) at high topgate bias and low temperature. These values are among the lowest reported for monolayer and bilayer $MoS_2$ so far.

The threshold backgate voltage $V_{bg}^{th}$, which is defined to be the bias voltage required to achieve a critical device current of 100 pA, shifts towards negative gate bias with increasing top-gate voltage as shown in Figure 2d. Note that for positive topgate voltages $V_{tg} > 0$ V the devices remain in the conducting state. It can also be seen that the shift of $V_{bg}^{th}$ is non-monotonous with an apparent change in the slope at around $V_{tg}$ = -0.6 V. We attribute this behavior to different capacitive coupling of the top and back gate with shift in $E_F$ and the corresponding changes in the quantum capacitance[33] $C_q$ of $MoS_2$. The total top gate capacitance between ion gel and $MoS_2$ channel $C_{tot}^{top}$ can be described as a series connection of $C_q$, which is proportional to the density of states (DOS) of the material, and the geometric capacitance of the ion gel $C_{ig}$. Similarly, the back gate capacitance $C_{tot}^{back}$ can be described as quantum capacitance and the oxide capacitance $C_{ox}$ in series. Ignoring screening effects, the top and back capacitances are given by:

$$C_{tot}^{top} = \left(\frac{1}{C_q} + \frac{1}{C_{ig}}\right)^{-1}$$

$$C_{tot}^{back} = \left(\frac{1}{C_q} + \frac{1}{C_{ox}}\right)^{-1}$$

Here, we can consider two regimes depending on the relative magnitude of the capacitances. When the channel is depleted and $E_F$ lies in the band gap, $C_q$ is very small due to the low density of mid-gap states[38]. This results in $C_q \ll C_{ig}$ and thus $C_{tot}^{top} \sim C_q$. Note that $C_{tot}^{back}$ is affected by the same effect, but to a lesser extent because $C_{ox}$ is intrinsically small. As a result, the ratio between $C_{tot}^{top}$ and $C_{tot}^{back}$ remains low. In the other limit, when the sample is strongly doped and $E_F$ is in the conduction band, $C_q$ significantly increases such that $C_{tot}^{top}$ has contributions from both $C_q$ and $C_{ig}$ while $C_{tot}^{back} \sim C_{ox}$. In this regime, the channel conductivity is much more efficiently tuned by the top gate bias as expected from the large capacitance of the ion gel. Thus, the ratio between $C_{tot}^{top}$ and $C_{tot}^{back}$ increases to a high value with increasing top gate voltage due to increase in $C_q$.

In the following, we discuss the high doping regime. Figures 3a-c show the color-coded map of the channel conductivity as a function of top and back gate bias. As indicated by the black dashed lines along constant conductivity, the top gate is 50 to 100 times more efficient compared to the back gate. The ratios $\Delta V_{bg}/\Delta V_{tg}$, where $\Delta V_{bg}$ and $\Delta V_{tg}$ denote the corresponding gate voltage difference to achieve the same change in conductivity in the metallic regime, are obtained from the slopes of the lines in Figure 3. This ratio is determined to be ~ 86 for monolayer, ~ 64 for bilayer and ~ 43 for trilayer device in the highly conductive region.

The difference in the gate coupling ratio for 1-3L samples can originate from various effects, such as the superposition of the two gate fields, changes of dielectric constants, different gate separation, and changes of $C_{ig}$ or $C_q$ (proportional to the DOS) as a function of layer thickness. Using a simple relationship $C_{tot}^{top}\Delta V_{tg} = C_{tot}^{back}\Delta V_{bg}$ (assuming $C_{tot}^{back} \sim C_{ox}$ for the case of frozen ions, see Hall measurements in the supporting information) from a capacitor model and assuming an ion gel capacitance[42] of $C_{ig} = 10.7$ µF/cm$^2$, we estimate the quantum capacitance of mono-, bi-, and trilayer samples to be 1.0 µF/cm$^2$, 0.8 µF/cm$^2$, and 0.5 µF/cm$^2$, respectively. These are rough estimates but are on the same order of magnitude as the theoretically prediction.[46] The trend contradicts the expectation that DOS increases with increasing thickness[47], suggesting that ion gel is effective only for the top layer which is in direct contact with the gel. Consequently, two parallel channels that are independently controlled may be present for few-layer systems. More accurate measurement of the quantum capacitance requires precise measurement of $C_{ig}$.

In order to show the full range of conductivity accessible by varying the two voltages, we show the back gate transfer curves displaced horizontally according to the top gate bias (Figure 4a and b). The insets show the transition from insulating to metallic conduction regimes where the temperature coefficient changes sign.[17] The crossover occurs at $V_{bg}$ ~ 80 V and $V_{tg}$ = 0 V for monolayer and $V_{bg}$ ~ 25 V and $V_{tg}$ = 0 V for bilayer. In accordance with previous results[17, 25, 26] the crossing points are at resistances on the order of h/e$^2$. Above the crossover point, conductivity decreases with increasing temperature, indicating phonon limited, metallic like transport. At lower charge carrier density, increasing conductivity with increasing temperature suggests thermally activated transport and conduction by variable range hopping.[13, 20, 22-24] As can be seen from figure 4b, ionic gating allows access to conduction regimes far beyond the crossover point where transport properties remains largely unexplored.

Field effect mobility and its dependence on carrier density and temperature offer insight into the fundamental transport properties of $MoS_2$.[48] The gate bias and temperature dependence of the field effect mobility ($\mu_{FE} = (1/C_{ox})*d\sigma/dV_{bg}$, where $C_{ox}$ = 11.5 nF/cm$^2$ is the back gate capacitance and $\sigma$ is the channel conductivity) for mono- and bilayer $MoS_2$ is shown in figures 4c-f. The mobility initially increases with the gate voltage, and eventually saturates to a constant value at higher gate biases. The mobility saturation can also be seen as a function of temperature below 30 K (Figure 4e and f). Note that the mobility becomes independent on both charge density and temperature in this regime. The saturation values are about 230 cm$^2$/Vs, 450 cm$^2$/Vs, and 820 cm$^2$/Vs for 1-3L devices, showing a rising trend with increasing layer numbers (see Supporting Information for 3L device data).

This mobility saturation agrees with earlier studies and indicates that at low temperatures and in the high charge density regime, charge transport in these devices is dominated by short-range or defect scattering.[20] Our results show the same behaviour not only in monolayer but also in bilayer and trilayer $MoS_2$ devices which indicates that intrinsic scattering mechanisms limit the device performance at low temperatures. It should be noted that despite the high concentration of ions, which are effectively charged impurities at the surface of the $MoS_2$, the low temperature mobility remains nearly constant with top gate bias. This observation suggests that the contribution of long-range scatterers (e.g. charged impurities) at the interface is less dominant compared to short-range scatterers (e.g. structural defects) in this regime.[17, 27] For temperatures above 100 K, the mobility decreases rapidly due to electron-phonon scattering and can be described by a power law dependence $\mu \sim T^{\gamma}$. Note that while the $\gamma$ factor seems to be almost constant for small variations of the charge carrier density with backgate voltage, using the topgate leads to very different values in the bilayer case from $\gamma = 1.94$ to 2.85.

In summary, we show that dual gating of mono- and few-layer $MoS_2$ using ion gel allows access to a wide range of carrier densities. We demonstrate low device resistivities of 1 kΩ and 420 Ω in mono- and bilayer devices at low temperatures, respectively. The analysis of the dual gating behavior of the devices reveals that the gate coupling is strongly dependent on the doping level and the number of layers. From the coupling ratios, we estimate the quantum capacitance to be on the order of 1 µF/cm$^2$ for all devices. Our transport results further shed light on the factors limiting carrier mobility in mono- and few layer $MoS_2$ in the highly conducting regime.

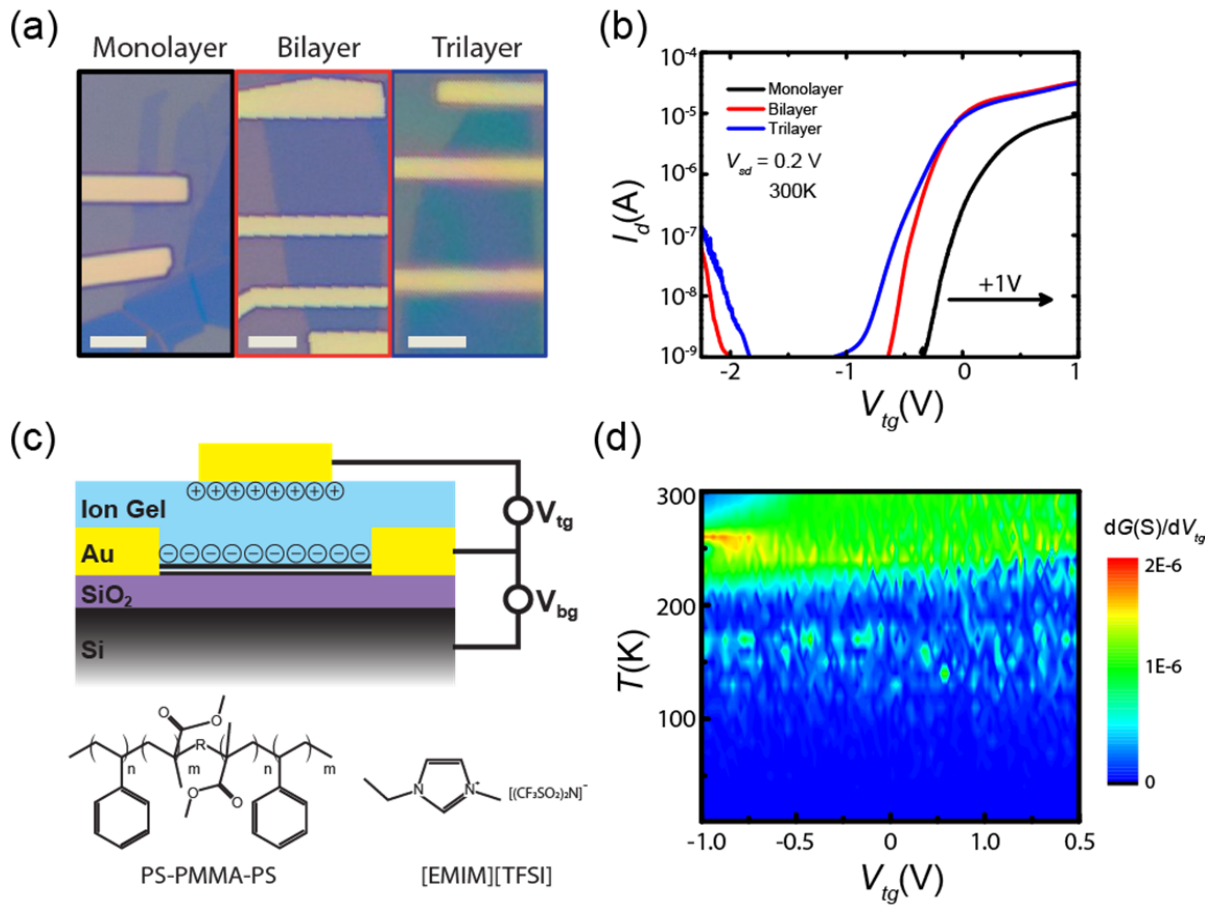

**Figure 1.** a) Optical images of mono-, bi-, and trilayer MoS$_2$ transistors covered with ion gel on silicon substrate, the scale bars indicate 5 μm. b) Ambipolar field-effect characteristics at room temperature. The curve for the monolayer sample is shifted by 1 V to the left. c) Schematic illustration of a dual gated transistor and chemical structures of the copolymer and the ionic liquid used as topgate. d) Transconductance of a bilayer device while sweeping $V_{tg}$ at fixed temperature. The blue part indicates the temperature regime, in which the ions are frozen out and top gate variations have no effect on the sample's conductance.

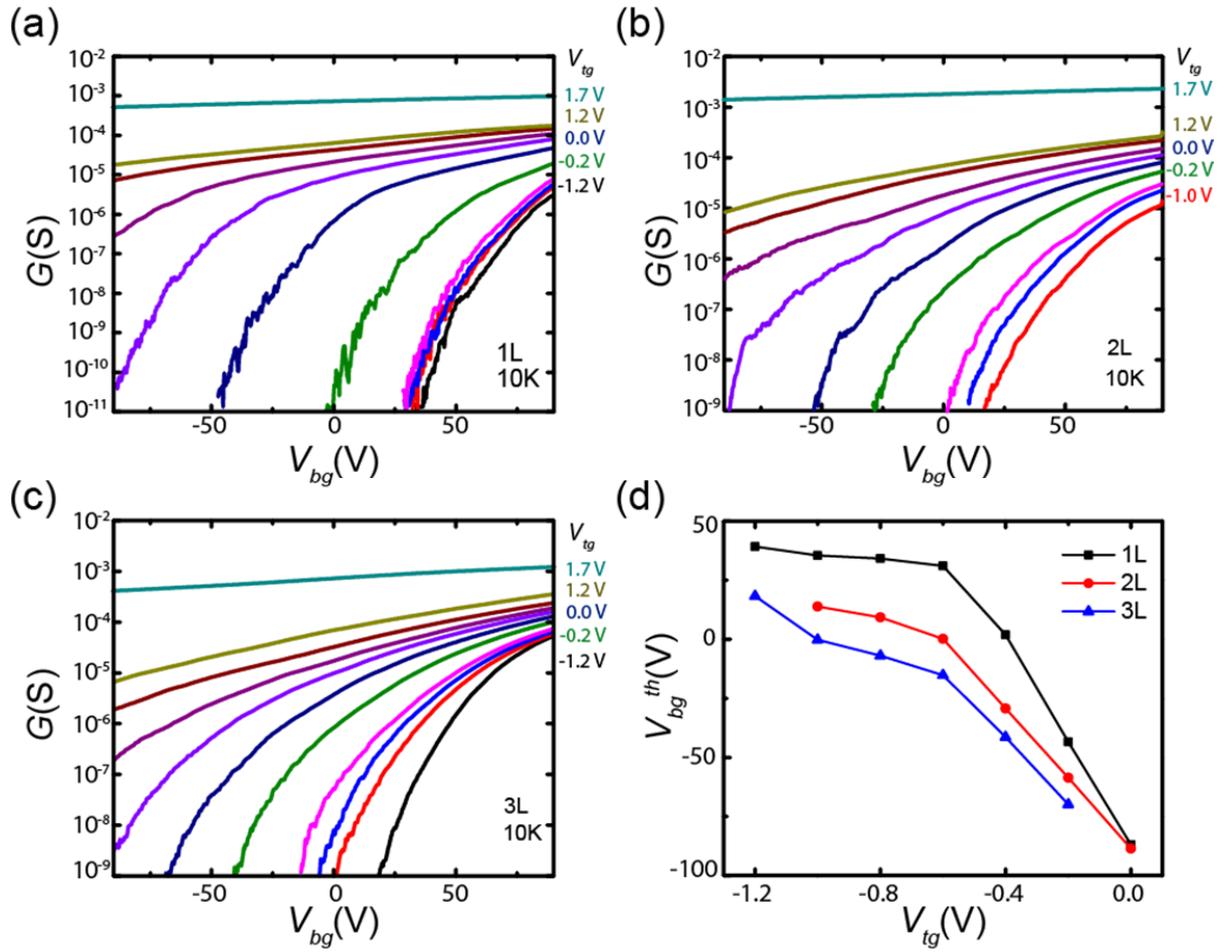

**Figure 2.** a-c) Channel conductivity as a function of $V_{bg}$ at different $V_{tg}$ (-1.2 V, -1.0 V, … 1.0 V, 1.2 V, 1.7 V) of 1-3L MoS$_2$ devices at 10 K. The minimal conductivity which can be reached is limited by the gate current and the measurement setup. d) The threshold voltage obtained from the cutoff in a-c) as a function of $V_{tg}$. For positive $V_{tg}$, $V_{bg}^{th}$ cannot be extracted since the sample is always in the on state in the range of backgate voltages studied.

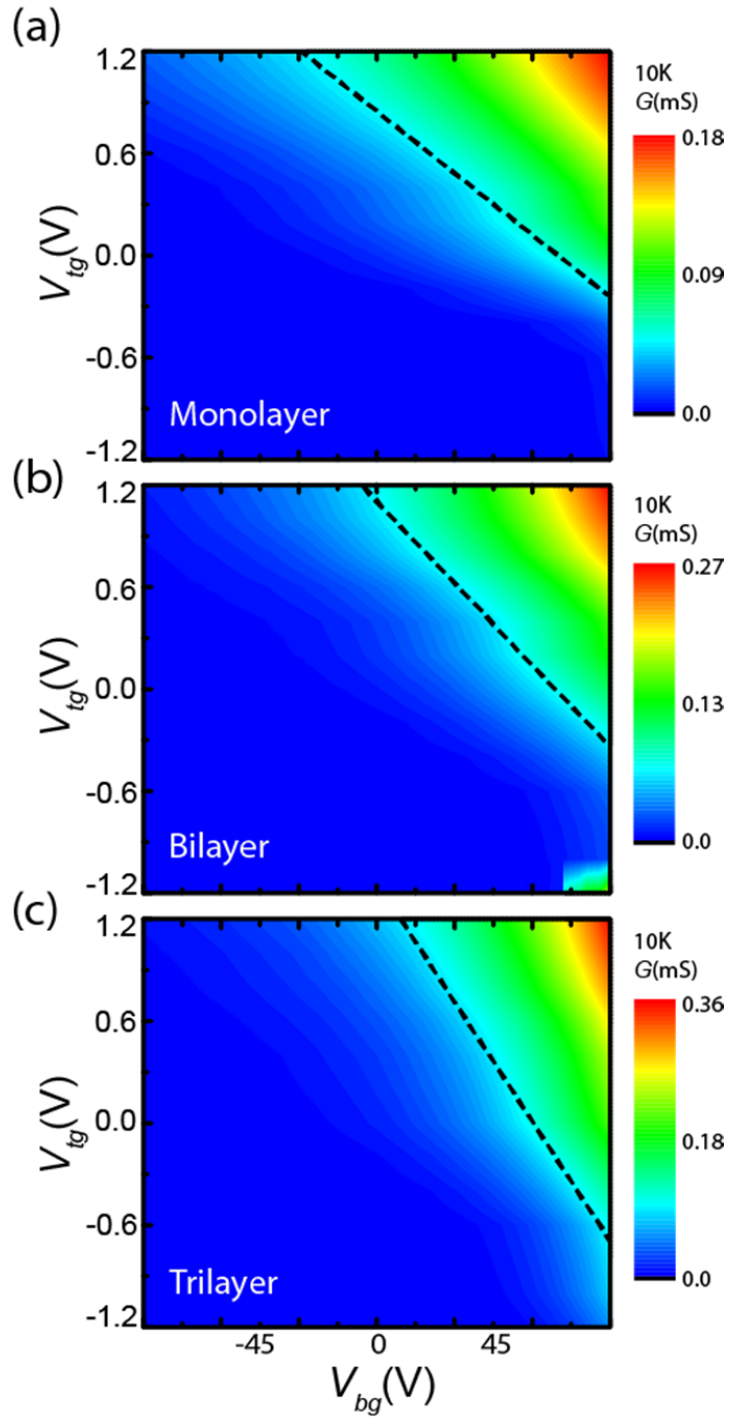

**Figure 3**. Color coded map of the dual gate characteristics for a) monolayer, b) bilayer and c) trilayer $MoS_2$. After setting the top gate voltage at room temperature and cooling the sample down to T=10 K, the backgate voltage has been varied.

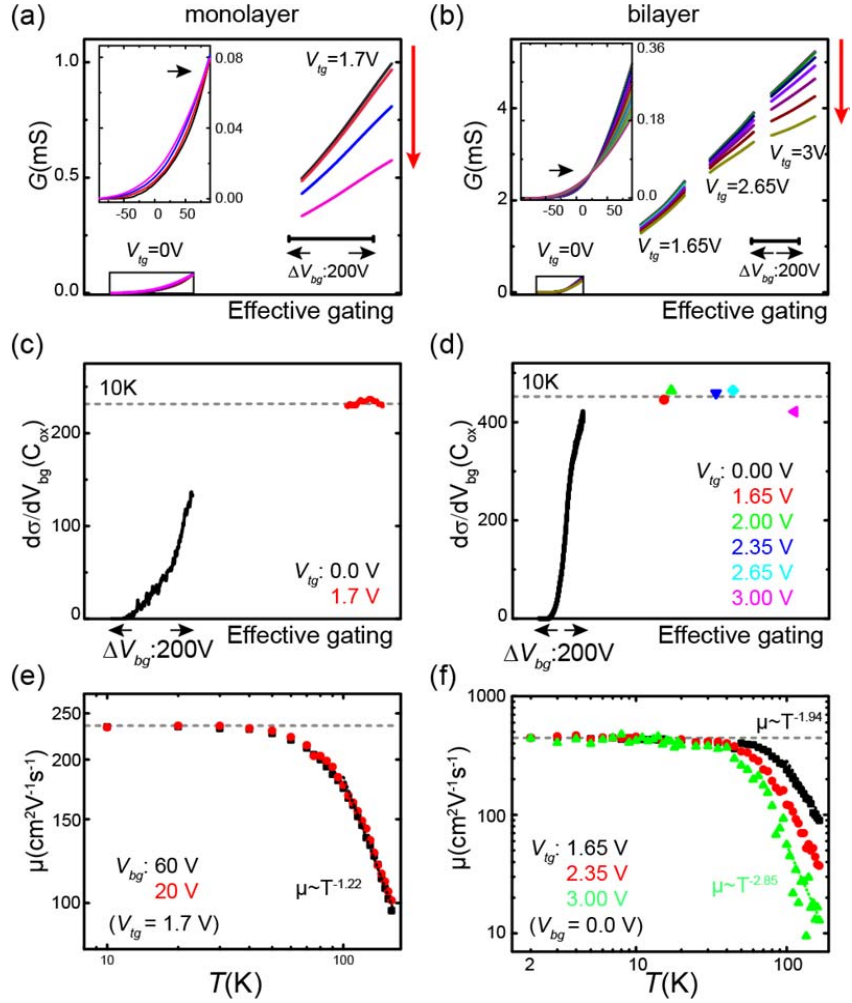

**Figure 4.** a,b) Transfer curves of monolayer and bilayer samples based on backgate sweeps at different top gate bias. The curves have been displaced horizontally as explained in the main text. The insets in a) and b) show the metal-insulator transition. The red arrows indicate increasing temperature from 10 K (3 K) to 150 K (60 K) for monolayer (bilayer) devices. c,d) Differential conductivity in units of the oxide capacitance at low temperature as a function of top and backgate voltage. For the monolayer device, we fix the top gate voltage at 1.7 V and vary the back gate voltage, while for the bilayer sample, the curves displayed are obtained at different top gate and 0 V back gate bias. e,f) Temperature dependence of measured mobility for monolayer and bilayer devices.

ACKNOWLEDGMENT

G.E acknowledges Singapore National Research Foundation for funding the research under NRF Research Fellowship (NRF-NRFF2011-02) and Graphene Research Centre. T.T. was partially supported by the Funding Program for the Next Generation of World-Leading Researchers and Grants-in-Aid from MEXT (26107533 "Science of Atomic Layers" and 25000003 "Specially Promoted Research"). B. Ö. acknowledges support by the Singapore Millenium Foundation-NUS Research Horizon award (R-144-001-271-592; R-144-001-271-646) and the NRF-CRP award (R-144-000-295-281).

ASSOCIATED CONTENT

**Supporting Information available:** Raman spectra for 1-3L samples, Hall measurements, band gap extraction, additional data on the 3L device. This material is available free of charge via the Internet at http://pubs.acs.org.

AUTHOR INFORMATION

**Corresponding Author**

#Email: g.eda@nus.edu.sg

**Author Contributions**

GE supervised the project, GE and HS designed the experiment, LC and SW prepared the samples, JP and TT supplied the ion gel and advised on the usage, LC and HS performed the measurements, GE, LC, and HS analysed the data, the manuscript was written through contributions of all authors and all authors have given approval to the final version of the manuscript.

ABBREVIATIONS
DOS, density of states; 2D, two-dimensional

Supporting information for

# Charge transport in ion-gated mono-, bi-, and trilayer MoS$_2$

*Leiqiang Chu\* [1,2], Hennrik Schmidt\* [1,2], Jiang Pu [4], Shunfeng Wang [1,2], Barbaros Özyilmaz [1,2], Taishi Takenobu [4,5,6], Goki Eda [1,2,3,#]*

[1]Graphene Research Centre, National University of Singapore, 6 Science Drive 2, Singapore 117546

[2]Department of Physics, National University of Singapore, 2 Science Drive 3, Singapore 117542

[3]Department of Chemistry, National University of Singapore, 3 Science Drive 3, Singapore 117543

[4]Department of Advanced Science and Engineering, Waseda University, Tokyo 169-8555, Japan

[5]Department of Applied Physics, Waseda University, Tokyo 169-8555, Japan

[6]Kagami Memorial Laboratory for Material Science and Technology, Waseda University, Tokyo 169-0051, Japan

\* These authors contributed equally to this work.

**Raman spectroscopy:**

To confirm the layer number of the flakes, Raman spectroscopy has been performed. The spectra are normalized to the intensity of the $A_{1g}$ peak. The peak separation between $E_{2g}^1$ and $A_{1g}$ can be used to determine the number of layers. As shown in Figure S1, the difference between the two peaks is found to be 18.3 cm$^{-1}$, 21.2 cm$^{-1}$, and 23.1 cm$^{-1}$ for monolayer (1L), bilayer (2L), and trilayer (3L) samples, respectively, consistent with previous reports.[1,2]

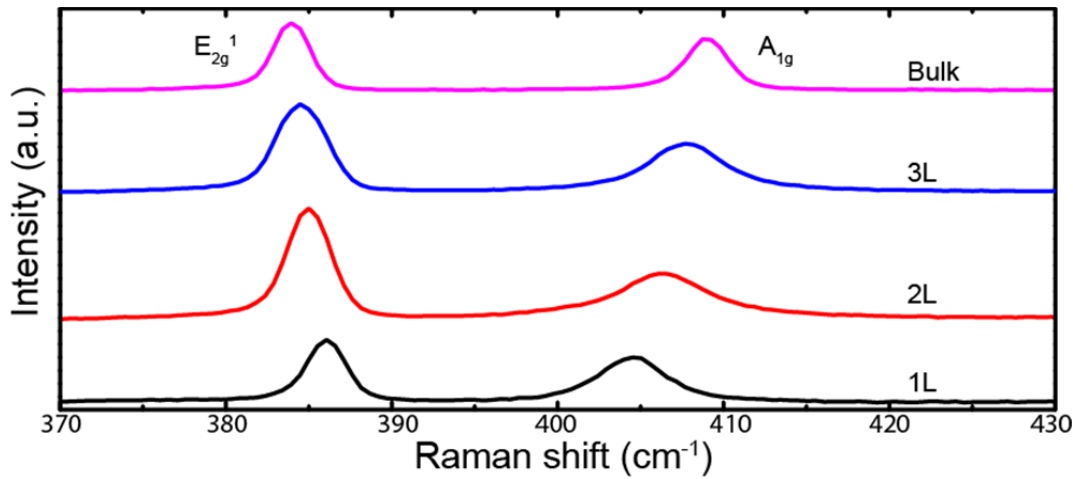

**Figure S1.** Normalized Raman spectra of mono-, bi-, and trilayer MoS$_2$ flakes as well as bulk.

**Capacitance of the back gate at low temperature:**

To extract low temperature mobility from the backgate transfer characteristics, we performed Hall measurement to study the capacitive coupling of top and back gate. Figure S2 shows charge carrier densities obtained from Hall measurements at a fixed topgate voltage and different backgate voltages at $T$=100 K. The change with backgate reassembles the linear behavior and slope which one would expect from a simple capacitance model for a device with the backgate only and indicates that the presents of a topgate with frozen ions has little effect on the backgate characteristics.

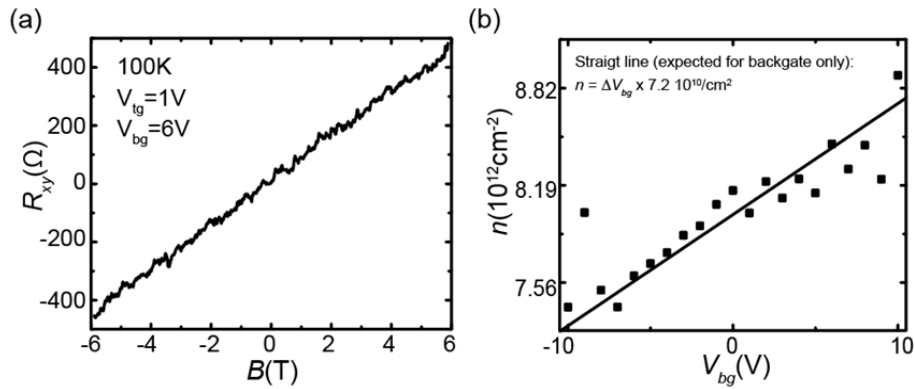

**Figure S2**. Hall measurement for a bilayer device. a) Hall resistance for $V_{tg} = 1$ V and $V_{bg} = 6$ V as a function of magnetic field at 100 K. b) Gate dependent carrier concentrations extracted from the Hall resistance. The straight line indicates the expected values without top gate covering.

**Band gap:**

As discussed in the main text, the quantum capacitance dominates when the Fermi level lies in the band gap, which enables the extraction of the band gap directly from the electron and hole threshold voltage.[3,4] As illustrated in Figure S3, the band gaps for bi- and trilayer samples are obtained to be approximately 1.63 eV and 1.32 eV.

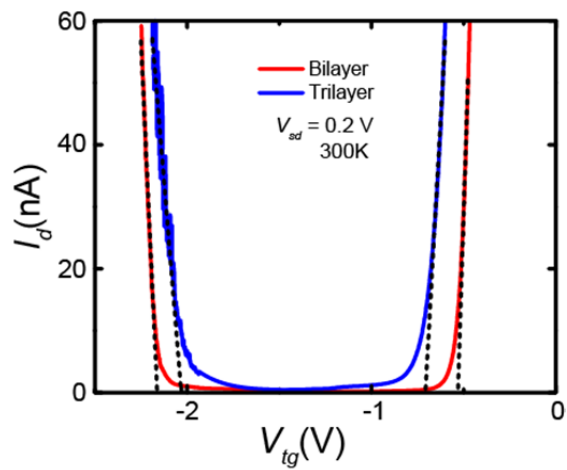

**Figure S3**. Extrapolation of the electron and hole branches of bilayer and trilayer devices to extract the threshold voltages, from which the band gap is determined.

### Further measurements on a trilayer device

Figure S4 shows the low temperature resistance of a trilayer sample as a function of top and bottom gate, displaced horizontally according to the top gate bias. In the panel below, the arrow indicates where the transition from insulating to metallic state happens.

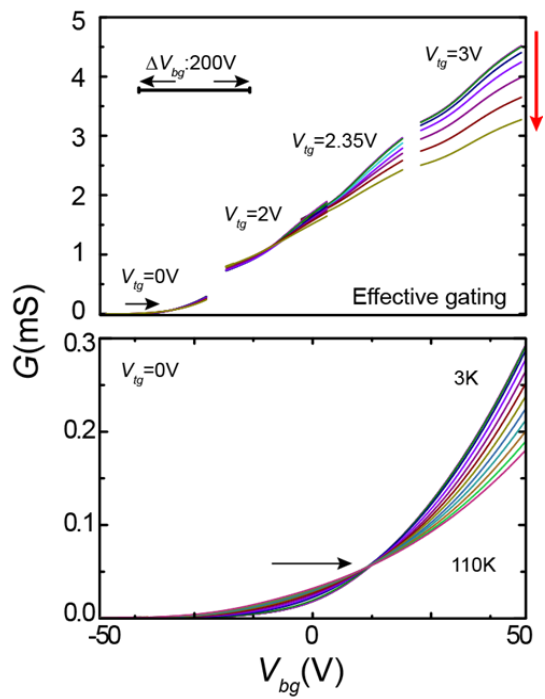

**Figure S4**. Transfer curves of trilayer sample at different top gate bias. The arrow indicates the metal-insulator transition in both panels. The red arrow indicates increasing temperature from 3 K to 60 K.

### Mobility of the trilayer device:

The gate bias and temperature dependence of the field effect mobility for trilayer $MoS_2$ is shown in figures S5a and b.

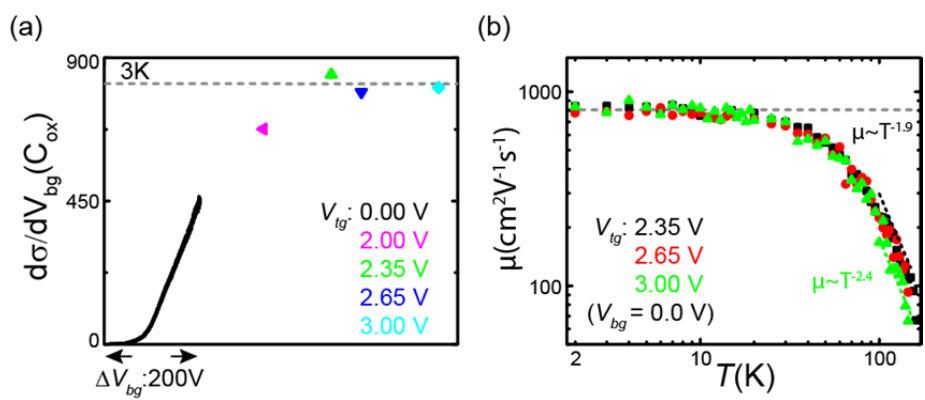

**Figure S5.** a) Differential conductivity in units of the oxide capacitance at 3 K as a function of top and backgate voltage for trilayer device. b) Temperature dependence of the measured mobility for the same device.

**Scattering mechanism:**

As shown in Figure 4 in the main text, above a temperature of 100 K mobility decreases rapidly due to electron-phonon scattering[5], which is often described by a power law dependence $\mu \sim T^{-\gamma}$. The damping factor $\gamma$ contains information on the type of phonons involved in the scattering.[6, 7] Simple fit of the raw experimental data yields damping factor ranging between 1.2 and 2.85 depending on the sample and doping level similar to earlier reports.[5]

One can subtract the contribution of short-range scattering from the observed mobility to obtain $\mu'$, which accounts for other scattering processes.[8] Figure S6 compares the as-measured mobility $\mu$ and extracted mobility $\mu'$ for all devices, respectively. It can be seen that $\mu'$ still follows the power law thus allowing extraction of a damping factor $\gamma'$. We find $\gamma'$ to be about 3.05, 3.0, and 2.54 for mono-, bi-, and trilayer devices, respectively.

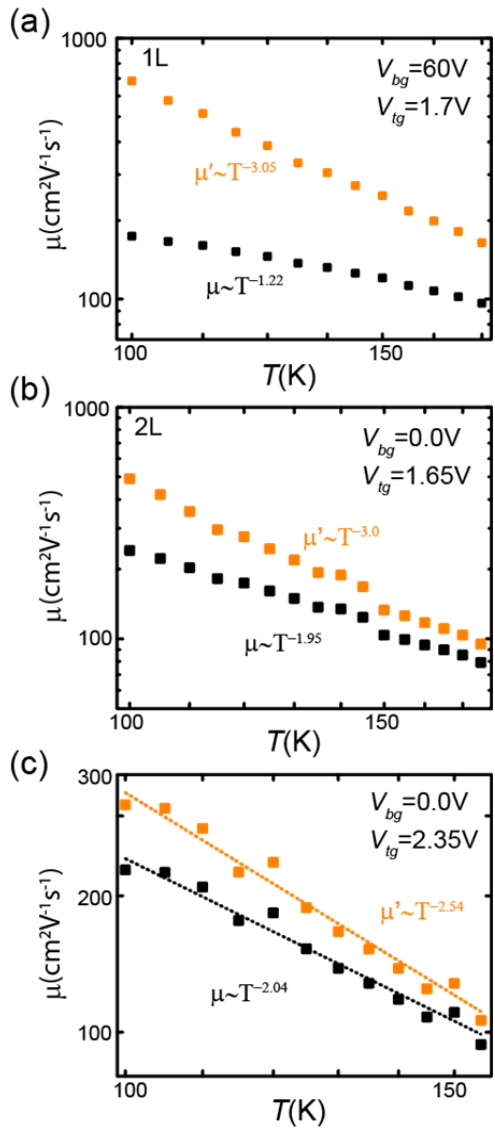

**Figure S6.** Temperature dependence of the as-measured mobility $\mu$ and extracted mobility $\mu'$ after subtracting the short range scattering part for dual gated monolayer (a), bilayer (b), and trilayer (c).